# Acoustomagnetoelectric effect in a degenerate semiconductor with nonparabolic energy dispersion law.

N. G. Mensah

Department of Mathematics and Statistics

University of Cape Coast, Ghana.

**Abstract:**

Surface (2D) acoustomagnetoelectric effect (SAME) and bulk (3D) acoustomagnetoelectric effect (AME) have been studied in a degenerate semiconductor with nonparabolic energy dispersion law. Expressions are obtained for the acoustomagnetoelectric fields i.e. $E_{SAME}$ and $E_{AME}$ under weak $\omega\tau \ll 1$ and strong $(\omega\tau \gg 1)$ magnetic fields. It is shown that $E_{SAME}$ and $E_{AME}$ depend on the nonparabolicity parameter $(\eta)$ in a complex manner. Unlike $E_{AME}$ which changes sign when the magnetic field **H** is reversed for $E_{SAME}$ this phenomenon occurs whenever $\theta$ satisfies the condition $\dfrac{\pi}{2} < \theta < \pi$.

**Introduction**

When acoustic phonons interact with conducting electrons, it leads to the absorption or amplification of acoustic phonons [1,2], acoustoelectric effect, (AE)[3–9], acoustomagnetoelectric effect (AME)[10-15], acoustothermal effect [16] and acoustoconcerntration effect.

Acoustomagnetoelectric effects which consists of generations of electrical current during the propagation of acoustic phonons through a conducting material placed in electrical and magnetic field, are defined in a small region in the spectrum of the phonon numbers. It can therefore be used in determining very vital information about the material under study, e.g. conductivity due to diffusion of charges, electromechanical constant, deformation potential and many others. The phenomenon also has immense applications in device production.

The AME was first predicted theoretically by Grinberg and Kramer [10] and observed experimentally in bismuth by Yamada [11]. Since then some works have done on it to understand the phenomenon. Epshtein and Gulyaev [12] studying this effect in a monopolar semiconductor noticed that AME effect occurs mainly because of the independence of the electron relaxation time on the energy, i.e. τ(ε) and that when τ=const the effect vanishes. In [12] Epshtein again attributed the scattering mechanism for the appearance of AME to the inelasticity of the electron scattering by optical phonons at low temperatures. The dependence of the AME on the anisotropy of the effective mass and relaxation time has also been reported in [18]. It is interesting to note that there are quite a number of experiments to investigate the effect [19 - 21]. In all these works the dispersion law was assumed to be quadratic.

The first work were Kane's nonpararbolic dispersion law was used can be found in [22]. In this paper the dependence of the AME field ($E_{AME}$) on the nonparabolicity parameter $\eta = \dfrac{T}{\varepsilon_g}$ is very different in weak (wτ<<1) and strong (wτ>>1) magnetic field. In weak fields the $E_{AME}$ decreases on increase of η whereas in strong field $E_{AME}$ increases with the increase of η.

Recent work on this effect by Kaganov et al. [13] found out that the effect is very sensitive to the structure of the electron spectrum. As a result it can even exist at τ=const. The effect has also been studied in the quantum regime by Galperin and Kagan [23] and Margulis and Margulis [14]. Mensah et al have studied the effect in semiconductor superlattice [15]. Their result is similar to that obtained by [14] for quantum acoustomagnetoelectric effect.

In this paper we shall consider AME in a degenerate semiconductor with nonparabolic energy dispersion law. We shall obtain the general expression of AME and then analysis of the result.

The paper is organized as follows: In section 1 we present the calculation and general solution and in Section 2 we discuss the results and draw some conclusions.

**CALCULATION**

We shall use Kane's model for nonstandard energy dispersion law which is given as

$$\varepsilon = \frac{\hbar^2 k^2}{2m_0} + \frac{\varepsilon_g}{2}\left[\left(1 + \frac{8k^2 p^2}{3\varepsilon_g^2}\right)^{1/2} - 1\right] \qquad (1)$$

When $\varepsilon_g >> kp$ i.e. if $k << 1$ we obtain the standard parabolic energy dispersion

$$\varepsilon = \frac{\hbar^2 k^2}{2m_0} + \frac{2p^2 k^2}{3\varepsilon_g} = \frac{\hbar^2 k^2}{2m_n} \qquad (2)$$

Where $\dfrac{1}{m_n} = \dfrac{1}{m_0} + \dfrac{4p^2}{3\hbar^2 \varepsilon_g}$  (3)

Cyclotron resonance experiment shows that the effective mass of electron at the bottom of the conduction band in InSb is several times smaller than the free electron mass, $m_n = 0.013 m_0$.

Hence in Eq(3) $\frac{1}{m_0}$ can be neglected and so $p$ becomes

$$p^2 = \frac{3\hbar^2 \varepsilon_g}{4m_n} \tag{4}$$

and Eq(1) becomes

$$\varepsilon = \frac{\varepsilon_g}{2}\left[\left(1+\frac{2\hbar^2 k^2}{m_n \varepsilon_g}\right)^{1/2} - 1\right] \tag{5}$$

From Eqs (4) and (5), for nonstandard energy dispersion law in the approximation $(m_0 >> m_n)$ $\varepsilon$ is written in terms of electron momentum $p$ as

$$\varepsilon = \frac{\varepsilon_g}{2}\left[\left(1+\frac{2p^2}{m_n \varepsilon_g}\right)^{1/2} - 1\right] \tag{6}$$

where m(ε) is given as

$$m = m(\varepsilon) = m_n\left(1+\frac{2\varepsilon}{\varepsilon_g}\right) \tag{7}$$

here $\varepsilon_g$ is the band gap and $m_n$ is the effective mass.

For surface acoustomagnetoelectric effect we shall consider the configuration where the acoustic phonon **w**, magnetic field **H** and the measured $E_{SAME}$ lie on the same plane. We shall further consider the situation where $ql >> 1$ ($q$ is the acoustic wavenumber and $l$ is the mean free path of an electron), and then solve the kinetic equation

$$-\left(eE + \omega_H[p,h], \frac{\partial f_p}{\partial p}\right) + \frac{p}{m}\frac{\partial f_p}{\partial r} =$$

$$-\frac{f_p - f_0(\varepsilon_p)}{\tau(\varepsilon_p)} + \frac{\pi \in^2, w}{\rho v_s^3}\{[f_{p+q} - f_p]\delta(\varepsilon_{p+q} - \varepsilon_p - \omega_q) + \quad (8)$$

$$-[f_{p-q} - f_p]\delta(\varepsilon_{p-q} - \varepsilon_p + \omega_p)\}$$

here $f_p$ is distribution function, $\rho$ is the density of the sample, $v_s$ is the velocity of sound, $\in$ is the constant of the deformation potential, $\tau(\varepsilon)$ is the dependence of relaxation time of electron on energy and $\omega_H = \frac{eH}{mc}$ (**e** is the electric charge, **H** is the magnetic field, **m** is the mass of electron as given in Eq(7) and **c** is the speed of light in vacuum). $\varepsilon$ is the energy of the electron as given in Eq(6). It is important to note that the units being used are such that $\hbar = 1$ and $k = 1$ where $\hbar$ (Planck's constant divided by 2π) and $k$ is Boltzmann's constant.

To solve Eq(8) we multiply it $by \frac{e}{m} p \, \delta(\varepsilon - \varepsilon_p)$ and sum it over **p.** We obtain for the partial current density $R(\varepsilon)$ the following expression.

$$\frac{R(\varepsilon)}{\tau(\varepsilon)} + \omega_H[h, R(\varepsilon)] = \Lambda(\varepsilon) + X(\varepsilon) \quad (9)$$

where

$$R(\varepsilon) = \frac{e}{m} \sum_p p f_p \delta(\varepsilon - \varepsilon_p) \tag{10}$$

$$\Lambda(\varepsilon) = -\frac{e}{m} \sum_p (E, \frac{\partial f_p}{\partial p}) p \delta(\varepsilon - \varepsilon_p) \tag{11}$$

$$X(\varepsilon) = \frac{\pi e \in^2 W}{\rho v_s^3 m} \sum_p p \delta(\varepsilon - \varepsilon_p) \{(f_{p+q} - f_p)$$

$$\delta(\varepsilon_{p+q} - \varepsilon_p - \omega_q) + (f_{p-q} - f_q) \delta(\varepsilon_{p-q} - \varepsilon_p + \omega_q) \} \tag{12}$$

In the linear approximation of $E$ and $W$ and considering $f_p$ as an equilibrium distribution function $f_0(p)$, we transform the summation into integrals and then integrate over the spherical coordinate to obtain the following expressions for $\Lambda(\varepsilon)$ and $X(\varepsilon)$.

$$\Lambda(\varepsilon) = -\frac{e^2 (2m_n)^{\frac{3}{2}}}{\rho v_s^3 m_n} \frac{\left(1 + \frac{\varepsilon}{\varepsilon_g}\right)^{\frac{3}{2}}}{\left(1 + 2\frac{\varepsilon}{\varepsilon_g}\right)} \varepsilon^{\frac{3}{2}} \frac{\partial f_0}{\partial \varepsilon} E \tag{13}$$

$$X(\varepsilon) = \frac{e \Gamma W}{m_n v_s} \frac{1}{f_0(\varepsilon)} \left(1 + \frac{2\varepsilon}{\varepsilon_g}\right) \frac{\partial f_0}{\partial \varepsilon} \vartheta(\varepsilon - \varepsilon_1) \tag{14}$$

where

$$\varepsilon_1 = \frac{\varepsilon_g}{2}\left\{\left(1+\frac{q^2}{2m_n\varepsilon_g}\right)^{1/2}-1\right\} \qquad (15)$$

and $\vartheta(\varepsilon-\varepsilon_1)$ - step function, and the coefficient of absorption of sound in this case becomes

$$\Gamma = \frac{\in^2 m_n q^2}{\pi\rho\, v_s} f_0(\varepsilon_1) \qquad (16)$$

Now solving Eq(8) with the help of Eq(13) and Eq(14) and considering the current density $j$ as

$$\vec{j} = \int_0^\infty R(\varepsilon)\, d\varepsilon \qquad (17)$$

we obtain

$$j = \frac{e^2 n}{m_n}\{\alpha_1 E + \omega_H \alpha_2[h,E] + \omega^2 \alpha_3 h(h,E)\} + \frac{e\Gamma}{m_n v_s}\{\beta_1 W + \omega_H \beta_2[h,W] + \omega_H^2 \beta_3 h(h,W)\} \qquad (18)$$

where $\omega_H = \frac{eH}{m_n c}$, $\alpha_k = <gu\tau^{k+l} v^{-k}>$, $\beta_k = <<g\tau^{k-1} v^{2-k}>>$

and $k = 1,2,3$

here $<...> = -\frac{(2m)^{1/2}}{3\pi^2 n_0} \int_0^\infty (...) \frac{\partial f}{\partial \varepsilon} \varepsilon^{1/2} \partial \varepsilon$ (19)

and

$$<<...>> = -\frac{1}{f_0(\varepsilon)} \int_0^\infty (...) \frac{\partial f_0}{\partial \varepsilon} \partial \varepsilon$$ (20)

$$u = \left(1 + \frac{\varepsilon}{\varepsilon_g}\right)^{1/2}; \quad v = \left(1 + \frac{2\varepsilon}{\varepsilon_g}\right); \quad g = \frac{\tau(\varepsilon)}{1 + \omega_H^2 \tau(\varepsilon)^2}$$

Considering a planer configuration where the acoustic phonon is directed along *X* axis and the magnetic field lying in the *XY* plane. The surface acoustomagnetic field $E_{SAME}$ will be paralled to the *Y* – axis. Assuming that the sample is open i.e. $j = 0$ solving Eq[18] will give $E_{SAME}$ as follows:

$$E_{SAME} = \frac{1}{2} E_w \sin 2\theta w_H^2 \left\{ \alpha_2 \beta_2 - \alpha_1 \beta_3 + \alpha_3 \beta_1 - \frac{\alpha_2^2 \beta_1}{\alpha_1} + \frac{\alpha_2}{\alpha_1}(\alpha_3 \beta_2 - \alpha_2 \beta_3) w_H^2 \right\}$$

$$\left\{ \alpha_1^2 + (\alpha_1 \alpha_3 + \alpha_2^2) w_H^2 + \frac{\alpha_2^2 \alpha_3}{\alpha_1} w_H^4 \right\}^{-1}$$ (21)

where $E_w$ is the acoustoelectric field.

**Discussion**

The complexity of Eq(21) makes analysis quite difficult. Therefore we shall consider special cases of it.

1. When the magnetic field is weak, i.e. $\omega_H \tau \ll 1$

    The expression for $E_{SAME}$ in Eq(21) reduces to

$$E_{SAME} = \frac{1}{2} E_W \sin 2\theta \, \omega_H^2 \{<u\tau v^{-2}><<\tau^2>>$$
$$-<u\tau v^{-1}><<\tau^3 v^{-1}>> + <u\tau^3 v^{-3}><<\tau v>> \qquad (22)$$
$$-<u\tau v^{-1}><u\tau^2 v^{-2}>^2 <<\tau v>>\}<uv^{-1}>^2$$

2. When the magnetic field is strong, i.e. $\omega_H \tau \gg 1$

$$E_{SAME} = \frac{1}{2} E_w \sin 2\theta \left\{ \frac{<<v^2>>}{<u>} - \frac{<<\tau v>>}{<u\tau v^{-1}>} \right\} \qquad (23)$$

From Eq(22) and Eq(23) it follows that in a weak magnetic field $E_{SAME}$ is proportional to $H^2$ whiles in a strong magnetic field it is independent of $H$. However when $\varepsilon_q \to \infty$ and the energy dispersion turns to parabolic energy dispersion law, Eq(22) and Eq(23) transforms to the result obtained in [24].

Further for degenerate semiconductor, where $\tau(\varepsilon)$ is given as

$$\tau(\varepsilon) = \tau_\circ \left(\frac{\varepsilon}{T}\right)^\gamma \frac{(1+ \varepsilon/\varepsilon_g)^\gamma}{1+2\varepsilon/\varepsilon_g} \qquad (24)$$

Inserting Eq(24) into Eq(22) and Eq(23) and solving after cumbersome manipulation, we obtained for weak field wτ<<1

$$E_{SAME} = \frac{1}{2} E_W \sin 2\theta \left(\frac{\mu H}{c}\right)^2 f_0^{-1}(0,\xi) F^0_{3/2,0}(\xi,\eta).$$

$$\cdot \left(F^0_{\gamma+\frac{3}{2},2}(\xi,\eta)\right)^{-2} \cdot \{F^0_{2\gamma+3/2,4}(\xi,\eta) F^0_{2\gamma,2}(\xi,\eta) -$$

$$- \left(F^0_{\gamma+\frac{3}{2},2}(\xi,\eta)\right)^{-1} \left(F^0_{2\gamma+3/2,4}(\xi,\eta)\right)^2 F^0_{\gamma,0}(\xi,\eta) +$$

$$F^0_{3\gamma+\frac{3}{2},6}(\xi,\eta) F^0_{\gamma,0}(\xi,\eta) - F^0_{\gamma+\frac{3}{2},2}(\xi,\eta) F^0_{3\gamma,4}(\xi,\eta)\} \quad (25)$$

where $\quad \mu = \dfrac{e\tau_0}{m}; f_0(x,y) = (e^{x-y}+1)^{-1}; x = \dfrac{\varepsilon}{T}$

$$\xi = \frac{\varepsilon_F}{T}; \eta = \frac{T}{\varepsilon_g};$$

$$F^v_{\gamma,k}(\xi,\eta) = \int_0^\infty \left(-\frac{\partial f_0}{\partial x}\right) \frac{x^v(x+\eta x^2)^v}{(1+2\eta x)^k} dx \quad (26)$$

Eq(26) is a Fermi's generalized Integral, values of it can be found in [25]

For strong magnetic field

$$E_{SAME} = \frac{1}{2} E_w \sin 2\theta f_0^{-1}(0,\xi) F^0_{\frac{3}{2},0}(\xi,\eta). \quad (27)$$

$$\cdot \left\{F^0_{0,-2}(\xi,\eta)\left(F^0_{3/2,0}(\xi,\eta)\right)^{-1} - F^0_{v,0}(\xi,\eta)\left(F^0_{v+\frac{3}{2},2}(\xi,\eta)\right)^{-1}\right\}$$

To compare the surface acoustomagnetoelectric effect 2D with bulk acoustomagnetoelectric effect 3D we solve Eq(18) using the following configuration

$\vec{H} \| OZ \quad \vec{W} \| OX, \quad E_{AME} \| OY$

We obtain for the weak magnetic field $\omega_H \ll 1$

$$E_{AME} = E_w \left( \frac{\mu H}{c} \right) f_0^{-1}(o,\xi) F_{\frac{3}{2},0}^0(\xi,\eta) \left( F_{\gamma+\frac{3}{2},2}^0(\xi,\eta) \right)^{-2} \cdot$$

$$\left\{ F_{2\gamma+\frac{3}{2},4}^0(\xi,\eta) F_{\gamma,0}^0(\xi,\eta) - F_{\gamma+\frac{3}{2},2}^0(\xi,\eta) F_{2\gamma,2}^0(\xi,\eta) \right\} \quad (28)$$

and for the strong field $\omega_H \tau \gg 1$

$$E_{AME} = E_w \left( \frac{\mu H}{c} \right)^{-1} f_0^{-1}(o,\xi) \left( F_{\frac{3}{2},0}^0(\xi,\eta) \right)^{-1}$$

$$\left\{ F_{\frac{3}{2},0}^0(\xi,\eta) F_{\gamma,4}^0(\xi,\eta) - F_{\nu-\frac{3}{2},2}^0(\xi,\mu) F_{0,-2}^0(\xi,\eta) \right\} \quad (29)$$

We observed that unlike $E_{AME}$ where change of sign occurs when the magnetic field is reversed, for $E_{SAME}$ change of sign occurs whenever it satisfies the condition $\frac{\pi}{2} < \theta < \pi$.

Finally for τ=constant Eq(28) and Eq(29) become

$$E_{AME} = E_\omega \left( \frac{\mu H}{c} \right) \left\{ \frac{<uv^{-2}><<v>> - <uv^{-1}>}{<uv>} \right\} \quad (30)$$

For $\omega_H \tau \gg 1$

and

$$E_{AME} = E_\omega \left(\frac{\mu H}{c}\right)^{-1} \left\{\frac{<u><<v>>}{<u>^2} - \frac{<u><v><<v>>}{<u>^2}\right\} \quad (31)$$

Eq (30) and (31) were first obtained in [22].

In conclusion we have studied acoustomagnetoelectric effect in the degenerate nonparabolic semiconductor for both 2D and 3D cases. We showed that $E_{SAME}$ and $E_{AME}$ depends on the nonparabolicity parameter $\eta$ in a very complex manner. Unlike $E_{AME}$ which changes sign when the magnetic field **H** is reversed for $E_{SAME}$, this phenomena occurs whenever $\theta$ satisfies the condition $\frac{\pi}{2} < \theta < \pi.$